\documentclass[english,prb,aps,twocolumn,preprintnumbers,amsmath,amssymb,array]{revtex4}
\usepackage[T1]{fontenc}
\usepackage{dcolumn}
\usepackage[latin1]{inputenc}
\usepackage{graphicx}
\usepackage{amssymb}
\usepackage{bm}

\makeatletter

\usepackage{dcolumn}
\usepackage{bm}

\usepackage{babel}
\makeatother

\begin{document}

\title{Negative Differential Spin Conductance by Population Switching}

\author{Guy Cohen$^{1}$ and Eran Rabani$^{2}$}

\affiliation{$^{1}$ School of Physics and Astronomy, The Sackler Faculty of Exact
Sciences, Tel Aviv University, Tel Aviv 69978, Israel\\
 $^{2}$ School of Chemistry, The Sackler Faculty of Exact Sciences,
Tel Aviv University, Tel Aviv 69978, Israel}

\date{\today}

\begin{abstract}
An examination of the properties of many-electron conduction through
spin-degenerate systems can lead to situations where increasing the
bias voltage applied to the system is predicted to decrease the
current flowing through it, for the electrons of a particular
spin. While this does not necessarily constitute negative differential
conductance (NDC) per se, it is an example of negative differential
conductance per spin (NDSC) which to our knowledge is discussed here
for the first time.  Within a many-body master equation approach which
accounts for charging effects in the Coulomb Blockade regime, we show
how this might occur.
\end{abstract} 
\maketitle

\section{introduction}
\label{sec:intro}
Nonequilibrium properties like electronic conduction in molecular
systems must be treated within a many-body nonequilibrium theory, and
the extensive body of effects such treatments produce has attracted
many research efforts on both the theoretical and experimental
sides.\cite{Dekker1999,Tour2000,Ratner2000,Joachim2000,Avouris2002,Nitzan2003_1,Heath2003,Datta2004,Hod2006a}
In particular, models of transport through molecules and quantum dots
have been shown to describe many nontrivial phenomena, of which the
Coulomb Blockade effect is a well known example.\cite{Elhassid00} Such
models have been shown to describe cases where an increase in the
source-drain bias on a small device coupled to macroscopic leads
actually results in a decrease of the current through
it.\cite{Ratner96,Lang97,Xue99,Heij99,Larade01,Rogge06} This nonlinear
behavior is known as negative differential resistance or conductance
(NDR or NDC), and its explanation must lie in the shifted states of
the system and the switching of electron populations between them, but
the exact mechanism may differ between the various cases.

Several such mechanisms, many of which are actually single-electron
effects, are worth noting: the resonant double-barrier tunneling
junction familiar in doped semiconductor work,\cite{Carroll63} where
an increasing bias pushes a resonant conduction state into the
conduction window and then out of the conduction band of one of the
electrodes, resulting in NDR;\cite{Larade01} the case in which the
electrodes themselves have narrow resonant features in their density
of states, like an atomic-scale STM tip or an atom weakly coupled to a
larger electrode, where the bias shifts the conducting levels of the
electrodes into and out of alignment with each
other;\cite{Avouris89,Lang97,Xue99} the Coulomb-Blockade case where
the bias charges the system in a way that kicks a level out of the
conduction window;\cite{Uozumi95,Shin98,Heij99} the more general case
where the biasing actually conforms the molecule or causes a change in
the interaction with phonons,\cite{Reed99,Nitzan04,Ratner04,Nitzan07}
resulting once again in fewer available conduction levels.

When the situation is complicated by the lifting of spin degeneracy,
the interplay between the occupation of spin levels and their coupling
to the leads can result in spin-dependent effects as well. This has
often been explored in cases where the leads are
ferromagnetic,\cite{Awschalom01,Zutic04} for example in the
spin-blockade or spin field effect transistor.\cite{Datta90} More
recently, spintronics without polarized leads have been
suggested,\cite{Takayanagi99,Richter01,D'Amico03,Richter03,Vasilopoulos04,Richter04a,Richter04b,Richter04c,Sigrist05,Vasilopoulos05,Peeters06,Perroni06,Cohen07}
where the leads would generally contain electrons of both spins, which
would scatter through the system with different transmission
properties. In a practical application the device might perform
various transformations on spins rather then just act as a current
switch.  Thus it makes sense to develop the concept of conduction or
resistance per spin, with the understanding that the same range of
nonlinear phenomenon that is of interest for the total current can
occur here for spin-dependent current. Specifically, the study of NDC
is naturally complemented by NDC per spin, which is exhibited whenever
increasing the bias voltage on a conducting device causes the current
through it \emph{for one spin} to decrease, while the total current
does not necessarily decrease.

In this paper we take an illustrative look at a novel mechanism for
the phenomenon of negative differential spin conduction (NDSC). We
show that in cases where the charging of a quantum dot is a dominant
energy scale of the problem and the spin degeneracy is lifted, NDSC
can occur. The basic mechanism involves population switching between
the two spin levels. In Section \ref{sec:model} we describe a simple
device in which NDSC may appear and be of interest, and explain the
multi-electron master equation approach we employ for calculating the
spin-polarized current. In Section \ref{sec:results} we display and
analyze the results. Finally, we discuss our conclusions in Section
\ref{sec:conclusions}.

Some of the topics touched upon in this work have also been addressed
by Raphy Levine over the past
decade.\cite{Levine02a,Levine02b,Levine02c,Levine03,Levine04,Levine05a,Levine05b}
It is a great honor to dedicate this work to him, on the occasion of
his $\mbox{70}^{th}$ birthday.

\section{Model}
\label{sec:model} 
Perhaps the simplest and most abstract spintronic device one might
imagine consists of a system with a single (energetically relevant)
electron level coupled to two metallic leads.  By making this level
non-degenerate in the spin degree of freedom in any desired way, one
can achieve filtering behavior by tuning the conduction window so as
to contain only one of the spin
levels.\cite{Takayanagi99,Richter03,Vasilopoulos04,Peeters06,Cohen07}
If the system is small, one expects that the charging should become an
important energy scale in the problem, and the population of electrons
the device contains at any given time should not vary greatly from the
neutral number of electrons. At this regime a many-electron master
equation
treatment~\cite{Beenakker91,Kinaret92,Bonet02,Hettler03,Datta2004,Elste05,Braig05,Mukamel06,Hanggi2006,Nitzan07}
can be expected to provide a good approximation of the dynamics,
particularly for larger voltages.

We consider a model for a spin-filter device which is described by a
two single spin levels. Such a device can be an atom, a molecule, a
quantum dot, or any system with discrete levels that are well
separated.  The single electron levels on the device are coupled to
two leads with chemical potentials $\mu_{L}$ and $\mu_{R}$ and
coupling constants $\gamma_{L(R),ij}\left(i=a,b\right)$, which we will
assume to be equal to $\gamma_{L(R),ij}=\gamma\delta_{ij}$. The model
Hamiltonian of such a device (not including the leads, since the
current and level populations will be calculated within a standard
multi-electron master equation approach)\cite{Datta2004} is given by:
\begin{equation}
H_{D}=\left(\varepsilon+\delta\right)\boldsymbol{a^{\dagger}a}+\left(\varepsilon-\delta\right)\boldsymbol{b^{\dagger}b}+\Delta\left(\boldsymbol{a^{\dagger}a}+\boldsymbol{b^{\dagger}b}-N_{0}\right)^{2},
\label{eq:hamiltonian}
\end{equation}
where $\boldsymbol{a}$ and $\boldsymbol{b}$ are single particle
annihilation operators corresponding to the two spin levels ($a \equiv
\uparrow$ and $b \equiv \downarrow$, respectively) and $N_{0}$ is the
neutral number of electrons.  We also introduce the spinless level
energy $\varepsilon$, the spin energy shift $\delta$ and the charging
energy $\Delta$. We note in passing that the notation is only for
convenience and no real assumptions are made as to the symmetry of the
shift. In fact, the results are relevant to any two separate channels
with different energies, for instance two quantum dots of slightly
different energies each of which is coupled to different leads with a
charging interaction between them.

We now describe the approach taken to construct the multi-electron
master equation, suitable for the above model, from single electron
data. If one neglects spin-dependent multi-electron effects, then it
is formally straightforward to build from a set of one-electron
Hamiltonian and spin eigenfunctions an anti-symmetric basis of
multi-electron wavefunctions. Limiting the discussion to only two
levels, one can define:
\begin{equation}
\Psi_{n_{1}n_{2}}=A_{12}\prod_{n_{i}=1}\varphi_{n_{i}}.
\label{eq:psi}
\end{equation}
Here $A_{12}$ is the two particle anti-symmetrization operator and the
states are identified by their (spin-dependent) level occupations
$n_{i}$ ($0$ or $1$ for fermions). Using this anti-symmetric
multi-electron wavefunction we can uniquely and conveniently determine
the nonzero matrix elements of a general many-body operator $G$
required to construct the master equation. According to the
Slater-Condon rules where only single electron integrals are taken
into account:
\begin{widetext}
\begin{subequations} \begin{eqnarray}
 &  & \langle\varphi_{i}|G|\varphi_{j}\rangle=g_{ij}\\
 &  & \langle\Psi_{n_{1}n_{2}}|G|\Psi_{n_{1}n_{2}}\rangle=\sum_{j=1}^{2}g_{jj}n_{j}\label{eq:sumrule}\\
 &  & \langle\Psi_{n_{1}n_{2}}|G|\Psi_{n'_{1}n'_{2}}\rangle=g_{11}\delta_{n_{2}n'_{2}}\delta_{n_{1},1-n'_{1}}+g_{22}\delta_{n_{1}n'_{1}}\delta_{n_{2},1-n'_{2}}\label{eq:diffrule}\\
    &  & \langle\Psi_{n_{1}n_{2}}|G|\Psi_{n'_{1}n'_{2}}\rangle=g_{12}\delta_{n_{2}-n'_{2}-1}\delta_{n_{1}-n'_{1}+1}+g_{21}\delta_{n_{2}-n'_{2}+1}\delta_{n_{1}-n'_{1}-1}.
\label{eq:crossrule}
\end{eqnarray}
\end{subequations}
\end{widetext}
Multi-electron effects will be considered only in the form of charging
energy. Since these values will be used in a rate-process calculation
rather than a full quantum formulation, constructing the
multi-electron states themselves is actually redundant, and
Eqs.~(\ref{eq:sumrule})-(\ref{eq:crossrule}) along with the single
particle data will provide all the necessary information.

The transfer rates between the multi-electron states are given
by:\cite{Datta2004}
\begin{equation}
R_{\ell,\alpha\rightarrow\beta}=\frac{\Gamma_{\ell,\alpha\beta}}{\hbar}Q_{\alpha\beta}^{\ell},
\label{eq:rate1}
\end{equation}
where the four multi-electronic states are labeled by the Greek
indices $|\alpha\rangle\equiv|n_{a}^{(\alpha)}n_{b}^{(\alpha)}\rangle$
and $|\beta\rangle\equiv|n_{a}^{(\beta)}n_{b}^{(\beta)}\rangle$, such
that (for instance) $|00\rangle$ is the empty state, $|01\rangle$ is
the state where only level $b$ is occupied, and $|10\rangle$ is the
state where only level $a$ is occupied. The lead index in the above is
$\ell\in\left\{ L,R\right\} $. For reasons that will become clear
below, we also define the total transfer rate summed over both leads:
\begin{equation}
R_{\alpha\rightarrow\beta}=\sum_{\ell}R_{\ell,\alpha\rightarrow\beta}.
\label{eq:rate2}
\end{equation}
Following the Slater-Condon rules (cf., Eqs.~(\ref{eq:diffrule}) and
(\ref{eq:crossrule})), the coupling between the multi-electron states,
$\Gamma_{\ell,\alpha\beta}$, is related to the single electron level
coupling (or the imaginary part of the self-energy)\cite{Datta_book}
and is given by $\Gamma_{\ell,\alpha\beta}=\gamma_{\ell,ii}(i=a,b)$ if
the two multi-electronic states differ only by the occupation of level
$i$, $\Gamma_{\ell,\alpha\beta}=\gamma_{\ell,ij}$ if they differ only
by $n_{i}$ and $n_{j}$ and $n_{i}-n_{j}=1$, and
$\Gamma_{\ell,\alpha\beta}=0$ otherwise. As noted above,
$\gamma_{\ell,ij}$ is the matrix element of the single electron level
coupling. $Q_{\alpha\beta}^{\ell}$ in Eq.~(\ref{eq:rate1}) is related
to the Fermi-Dirac function, $f(\epsilon)$:
\begin{equation}
Q_{\alpha\beta}^{\ell}=
\begin{cases}
f(\epsilon_{\alpha}-\epsilon_{\beta}-\mu_{\ell}) &
N_{\alpha}>N_{\beta},\\
1-f(\epsilon_{\alpha}-\epsilon_{\beta}-\mu_{\ell}) &
N_{\alpha}<N_{\beta},\\ 1 &
N_{\alpha}=N_{\beta},
\end{cases}
\label{eq:Q}
\end{equation}
where $N_{\alpha}=\sum_{i}n_{i}^{(\alpha)}$ is the number of electrons
in state $\alpha$. The state energies are calculated from the
Hamiltonian~(\ref{eq:hamiltonian}) and amount to $\Delta N_{0}^{2}$,
$\left(\varepsilon+\delta\right)+\Delta\left(1-N_{0}\right)^{2}$,
$\left(\varepsilon-\delta\right)+\Delta\left(1-N_{0}\right)^{2}$ and
$2\varepsilon+\Delta\left(2-N_{0}\right)^{2}$ respectively for the
states $|00\rangle$, $|10\rangle$, $|01\rangle$ and $|11\rangle$.

Once the rates are known, the linear master equation system can be
read from the detailed-balance condition for steady-state:
\begin{equation}
\sum_{\beta}R_{\alpha\rightarrow\beta}P_{\alpha}-\sum_{\beta}R_{\beta\rightarrow\alpha}P_{\beta}=0,
\label{eq:master}
\end{equation}
where $P_{\alpha}$ is the probability that the system is in a
multi-electron state $\alpha$. The current at steady state is given in
terms of the steady state occupation probabilities and can be
expressed as~\cite{Datta2004}:
\begin{equation}
I_{\ell}=-e\sum_{\alpha\beta}R_{\ell,\alpha\rightarrow\beta}P_{\alpha}s_{\alpha\beta},
\label{eq:current1}
\end{equation}
where
\begin{equation}
s_{\alpha\beta}=
\begin{cases}
+1 & N_{\alpha}<N_{\beta},\\
-1 & N_{\alpha}>N_{\beta},\\
0 & N_{\alpha}=N_{\beta}.
\end{cases}
\label{eq:s1}
 \end{equation}
Intuitively, this expression states that current flows out of lead
$\ell$ whenever an electron flows from it into the device, with the
inverse also true. Following a similar line of physical reasoning
leads to an expression for spin-polarized current: up or down current
flows out of lead $\ell$ whenever an up or down electron flows from it
into the device. Assuming no coupling between levels with different
spin, the spin-dependent current is given by:
\begin{equation}
I_{\ell,a(b)}=-e\sum_{\alpha\beta}R_{\ell,\alpha\rightarrow\beta}P_{\alpha}s_{
a(b)\alpha\beta},
\label{eq:current2}
\end{equation}
and
\begin{equation} s_{a(b)\alpha\beta}=
\begin{cases}
+1(0) & S_{\alpha}<S_{\beta}\wedge N_{\alpha}<N_{\beta},\\ 0(+1) &
S_{\alpha}>S_{\beta}\wedge N_{\alpha}<N_{\beta},\\ 0(-1) &
S_{\alpha}<S_{\beta}\wedge N_{\alpha}>N_{\beta},\\ -1(0) &
S_{\alpha}>S_{\beta}\wedge N_{\alpha}>N_{\beta}.
\end{cases}
\label{eq:s2}
\end{equation}
Here, $S_{\alpha}=\sum_{i}s_{i}^{(\alpha)}$ where
$s_{i}^{(\alpha)}=\pm1$ for spin up ($a$) or down ($b$), respectively.

The linear master equations can be solved analytically, but the form
of the solution is rather cumbersome and has no real benefit. They can
also, of course, be solved numerically, which is the method chosen for
this work.

\section{Results and Analysis}
\label{sec:results}
\begin{figure}
\includegraphics[width=8cm]{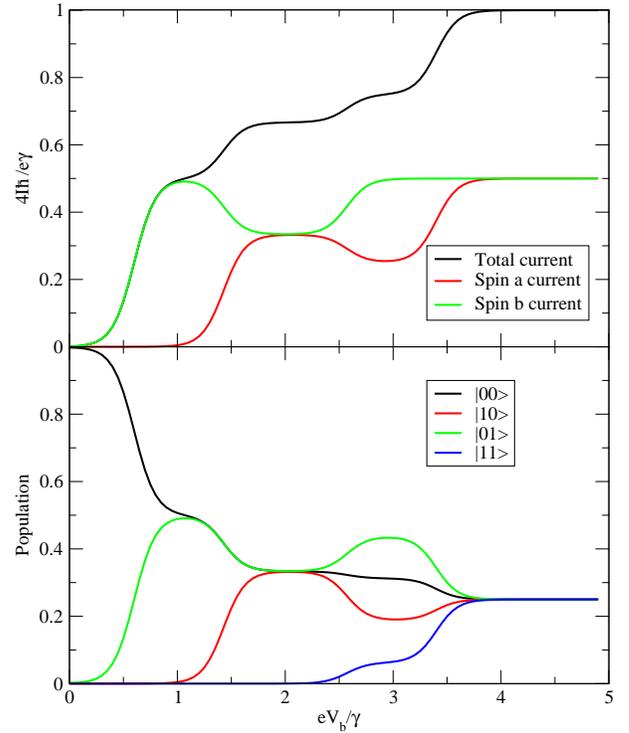}
\caption{The current (upper panel) and state populations (lower panel)
as a function of bias voltage, for the parameters $\beta=20/\gamma$,
$\mu=0$, $\epsilon=\gamma$, $\delta=\gamma/5$ and $\Delta=\gamma/2$,
where the neutral number of conduction electrons has been taken to be
$N_{0}=1$ and the voltage is applied symmetrically. Note the distinct
region of negative differential conduction (decreasing current) for
each spin current, which always occurs simultaneously with a decrease
in population for a state populated with the matching spin.}
\label{fig:ME-filter}
\end{figure}

As the above model is more or less identical to the one commonly used
to explain the Coulomb blockade, it is no surprise that plotting the
$I-V$ characteristics of the system shown in the upper panel of
Fig.~\ref{fig:ME-filter} immediately displays the well-known nonlinear
current steps typical of Coulomb blockade effect.\cite{Elhassid00} The
model parameters used here are $\beta=20/\gamma$, $\mu=0$,
$\epsilon=\gamma$, $\delta=\gamma/5$ and $\Delta=\gamma/2$. The
neutral number of conduction electrons has been taken to be
$N_{0}=1$. Conduction peaks or rises in the total current are expected
in this formalism when there exists an energy difference between two
states with electronic occupations that differ by one
($N_{\alpha}-N_{\beta}=\pm1$), which is also the energy of an electron
occupied in one lead but not the other (when $\Delta E\equiv
E_{\alpha}-E_{\beta}=\mu\pm eV/2$).  In other words, the total current
can rise at any bias voltage $V_{B}$ where the conduction window is
expanding so as to contain some spectral line of the system.  For the
present model, this occurs when
\begin{equation}
\left|eV_{B}\right|=-\mu\pm
\begin{cases}
\varepsilon\pm\delta+\Delta\left(\left(1-N_{0}\right)^{2}-N_{0}^{2}\right),\\
\varepsilon\pm\delta+\Delta\left(\left(2-N_{0}\right)^{2}-\left(1-N_{0}\right)^{2}\right).
\end{cases}
\label{eq:peak-locs}
\end{equation}
This is clearly the case for the current shown in the upper panel of
Fig.~\ref{fig:ME-filter}, where the four steps observed in the total
current appear at $0.6\gamma$, $1.4\gamma$, $2.6\gamma$, and
$3.4\gamma$ corresponding to transitions between states
$|00\rangle\leftrightarrow|01\rangle$,
$|00\rangle\leftrightarrow|10\rangle$,
$|01\rangle\leftrightarrow|11\rangle$, and
$|10\rangle\leftrightarrow|11\rangle$ , respectively and according to
Eq.~(\ref{eq:peak-locs}).

Turning now to discuss the current per spin also shown in the upper
panel of Fig.~\ref{fig:ME-filter}, we still observe the Coulomb
blockade steps, however, the direction of the step can be either
positive or negative. This is an example of a negative differential
spin conduction where an increase of the bias voltage is followed by a
decrease in the current per spin. The NDSC occurs for both spins in
this case, and at a different bias voltage for each spin-current. The
first drop in current occurring for spin type $b$ is also accompanied
by a sudden drop in the population of the $|01\rangle$ state (shown in
the lower panel of Fig.~\ref{fig:ME-filter}), as the change in
chemical potentials begins to allow the population of the $|10\rangle$
state. This switching of populations between the states is reminiscent
of another example of nonmonotonic changes in occupation predicted to
occur in a system of two electrostatically coupled single-level
quantum dots.\cite{Gefen05,Oreg05}.

There is a simple ``hand waving'' explanation for such behavior: the
current for each spin consists at low bias of contributions
proportional to the probability that the system is in some state
$\alpha$ and to the rate of transitions between state $\alpha$ and
state $|00\rangle$ (in which none of the states are occupied), where
$\alpha$ can be either $|10\rangle$ for spin up current or
$|01\rangle$ for spin down current. We therefore expect that at any
chemical potentials where the relevant Fermi functions and hence the
rates are nearly constant at the relevant energy, the current will, to
a good approximation, be linearly proportional to the population. The
population, in turn, decreases whenever the shifted energetics allow
the occupation of a new state. For this to be possible, the bias must
be applied in such a way that not all states become occupied
simultaneously. This is the reason the central chemical potential has
been placed below the levels.The second current drop in the figure,
which occurs at higher voltage and for the $a$ spin current, can be
explained by a similar argument - this time, however, the depopulation
of the $|01 \rangle$ state is the one involved.

It is worth pointing out that the population shifts are such that in
regions where the chemical potentials are far from any levels, any
states that energetically can be populated become so with equal
probability.  At low bias only one state has the entire population,
then as the bias is increased the population is shared equally between
two states, then between three, and finally between all four states.
This is the cause of the downward shifts in the population which
result in the NDSC.  For the example shown in Fig.~\ref{fig:ME-filter}
conduction sets in when the population of state $|00\rangle$ decreases
from its maximal value of $1$ until both states $|00\rangle$ and
$|01\rangle$ are equally populated.  Then NDSC occurs when both states
$|00\rangle$ and $|01\rangle$ lose population to state $|10\rangle$
until all three state become equally populated.  It is also evident that
in systems with more electronic levels, NDSC due to population
switching will become weaker if the separation between the states is
small.  Noticing this fact also clarifies the role of charging in
NDSC, as without charging all the states which include the same
energetically occupiable levels would become populated at the same
bias voltage.

\begin{figure}
\includegraphics[width=8cm]{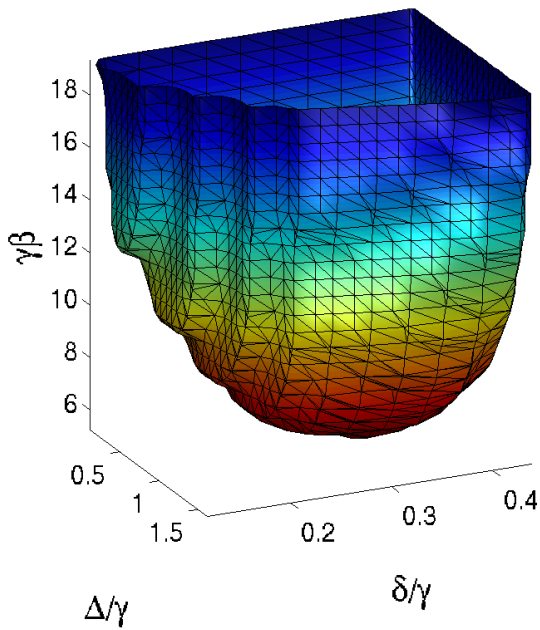}
\caption{A plot of a part of the surface of transition between regions
where NDSC does and does not occur for spin $b$ is shown, for
$\epsilon=\gamma$ and $\mu=0$, with the voltage applied
symmetrically. Note that at the limits of high temperature, small
charging energy and small level splitting there is no NDSC, but that
in general the behavior is complex.}
\label{fig:transition-surface}
\end{figure}

The theoretical phenomenon of NDSC and its physics are easy to
understand, and the mechanism we suggest for it here simple, but two
important questions remain: when will it occur, and how can it be
observed experimentally?  Answering the first question formally is a
matter for the analysis of the expressions for the spin currents: by
taking their derivatives and looking for a local maximum in the
voltage, an exact condition could be worked out in principle. In
practice, the analytical development involves the solution of
nonlinear equations, may or may not be possible and of interest, and
is beyond the scope of the present work. Instead, a look at a part of
the surface of transition in parameter space between regions where
NDSC does and does not appear (see Fig.~\ref{fig:transition-surface})
is enough to convince oneself that the exact conditions for NDSC are
nontrivial. If one is more interested in the approximate limits where
the drop in the current constitutes a sizable fraction of its maximum
value and where the master equation is valid, these can be expected
when the temperature is smaller than the splitting between the spin
states, $\beta \delta \gtrsim 1$; when the conduction resonances are
narrow enough such that $\frac{1}{\beta},\delta \gtrsim \gamma$; and
when the charging energy is of the order of the level spacing,
$\Delta\sim\epsilon\gtrsim\gamma$.  All criteria can be met, for
example, for systems of nanometer dimensions where the charging energy
and the level spacing can be tuned by simple changing the size. Also,
as mentioned above, some asymmetry in the application of the chemical
potential and/or bias voltage is required, as the spin dependent
effects happen to cancel out completely when the chemical potential is
exactly between the energies of the two levels and the bias is applied
symmetrically. In addition, NDSC requires some state to become
energetically occupiable at a bias voltage higher than one at which a
spin current exists. More precise conclusions require a calculation
similar to the one done to produce Fig.~\ref{fig:transition-surface},
which takes negligible computational effort and can be easily extended
to more detailed scenarios. However, the effect clearly occurs for an
extraordinarily wide range of parameters, as can be seen in the
figure.

Addressing the second question posed above, pertaining to experimental
observability, the following is proposed: NDSC is obviously equivalent
to NDC whenever the spin components of the current are observed
separately.  This can be achieved in any experiment where in addition
to flowing through the system described here, the current also flows
(while retaining spin coherence) through some sort of spin beam
splitting device which separates it into spin components before the
current is measured.  Such devices have been suggested in previous
works.\cite{Peeters06,Cohen07}

\section{Summary and Conclusions}
\label{sec:conclusions}
Using the well-established calculational methodology of multi-electron
master rate equations and a simple model of a quantum dot coupled to
metallic leads, we have pointed out a mechanism that gives rise to
negative differential spin conduction. The fact that NDSC occurs in
such a basic model for a wide range of parameters suggests that it
represents a real physical phenomenon. We have also discussed when
effects of this type can be expected to occur (the temperature must be
low enough, conduction peaks narrow, and charging should be
significant as this is strictly a many-particle effect), and suggested
how an experiment in which they might be measured could be carried
out.

The NDSC effect, as caused by population switching or any other
mechanism, is similar to and in special circumstances identical to the
NDC effect which has been observed in a variety of nano- and
meso-scale experiments.  It is characterized by an increase in bias
voltage over a junction resulting in the decrease of the current for
electrons of one particular spin. One way to observe NDSC directly is
to measure the current after it passes through a beam
splitter. Regarding the population switching mechanism: if the
energetics are tuned so that increasing the bias allows the sequential
occupation of several states, which with charging limiting the total
occupation results in a nonmonotonic behavior of the populations, NDSC
can be caused by a decrease in the population of a state which is
instrumental in the conduction of one spin. When this happens without
significantly effecting the electronic flow rates to and from that
state, a decrease occurs in the contribution to the current from the
term proportional to the population, leading to NDSC.

\section{Acknowledgments}
\label{sec:acknowledgments}
We would like to thank Roi Baer, Yuval Gefen, Oded Hod, Andrew Millis,
Abe Nitzan, Yuval Oreg, and Yoram Selzer, for discussions and
suggestions. This work was supported by the Israel Ministry of
Science.


\begin{thebibliography}{10}

\bibitem{Dekker1999}
C.~Dekker, {\em Physics Today\/} {\bf 52}, 22 (1999).

\bibitem{Tour2000}
J.~M. Tour, {\em Acc. Chem. Res.\/} {\bf 33}, 791 (2000).

\bibitem{Ratner2000}
M.~Ratner, {\em Nature\/} {\bf 404}, 137 (2000).

\bibitem{Joachim2000}
C.~Joachim, J.~k.~Gimzewski, and A.~Aviram, {\em Nature\/} {\bf 408}, 541
  (2000).

\bibitem{Avouris2002}
P.~Avouris, {\em Acc. Chem. Res.\/} {\bf 35}, 1026 (2002).

\bibitem{Nitzan2003_1}
A.~Nitzan and M.~A. Ratner, {\em Science\/} {\bf 300}, 1384 (2003).

\bibitem{Heath2003}
J.~R. Heath and M.~A. Ratner, {\em Physics Today\/} {\bf 56}, 43 (2003).

\bibitem{Datta2004}
S.~Datta, {\em Nanotechnology\/} {\bf 15}, S433 (2004).

\bibitem{Hod2006a}
O.~Hod, E.~Rabani, and R.~Baer, {\em Acc. Chem. Res.\/} {\bf 39}, 109 (2006).

\bibitem{Elhassid00}
Y.~Elhassid, {\em Rev. Mod. Phys.\/} {\bf 72}, 895 (2000).

\bibitem{Ratner96}
V.~Mujica, M.~Kemp, A.~Roitberg, and M.~A. Ratner, {\em J. Chem. Phys.\/} {\bf
  104}, 7296 (1996).

\bibitem{Lang97}
N.~D. Lang, {\em Phys. Rev. B\/} {\bf 55}, 9364 (1997).

\bibitem{Xue99}
Y.~Xue, S.~Datta, S.~Hong, R.~Reifenberger, J.~I. Henderson, and C.~P. Kubiak,
  {\em Phys. Rev. B\/} {\bf 59}, 7852 (1999).

\bibitem{Heij99}
C.~P. Heij, D.~C. Dixon, P.~Hadley, and J.~E. Mooij, {\em Appl. Phys. Letts.\/}
  {\bf 74}, 1042 (1999).

\bibitem{Larade01}
B.~Larade, J.~Taylor, H.~Mehrez, and H.~Guo, {\em Phys. Rev. B\/} {\bf 64},
  075420 (2001).

\bibitem{Rogge06}
M.~C. Rogge, F.~Cavaliere, M.~Sassetti, R.~J. Haug, and B.~Kramer, {\em New J.
  of Phys.\/} {\bf 8}, 298 (2006).

\bibitem{Carroll63}
J.~M. Carroll, editor, {\em Tunnel-Diode and Semiconductor Cicuits\/}
  (McGraw-Hill, New York, 1963).

\bibitem{Avouris89}
I.-W. Lyo and P.~Avouris, {\em Science\/} {\bf 245}, 1369 (1989).

\bibitem{Uozumi95}
H.~Nakashima and K.~Uozumi, {\em J. Appl. Phys.\/} {\bf 34}, L1659 (1997).

\bibitem{Shin98}
M.~Shin, S.~Lee, K.~W. Park, and E.~H. Lee, {\em Phys. Rev. Lett.\/} {\bf 80},
  5774 (1998).

\bibitem{Reed99}
J.~Chen, M.~A. Reed, and A.~M. R. J.~M. Tour, {\em Science\/} {\bf 286}, 1550
  (1999).

\bibitem{Nitzan04}
M.~Galperin, M.~A. Ratner, and A.~Nitzan, {\em Nano Lett.\/} {\bf 5}, 125
  (2004).

\bibitem{Ratner04}
A.~Troisi and M.~A. Ratner, {\em Nano. Lett.\/} {\bf 4}, 591 (2004).

\bibitem{Nitzan07}
M.~Galperin, M.~A. Ratner, and A.~Nitzan, {\em J. Phys. Cond. Matt.\/} {\bf
  19}, 103201 (2007).

\bibitem{Awschalom01}
S.~A. Wolf, D.~D. Awschalom, R.~A. Buhrman, J.~M. Daughton, S.~{von Molnar},
  M.~L. Roukes, A.~Y. Chtchelkanova, and D.~M. Treger, {\em Sceince\/} {\bf
  294}, 1488 (2001).

\bibitem{Zutic04}
I.~Zutic, J.~Fabian, and S.~{Das Sarma}, {\em Rev. Mod. Phys.\/} {\bf 76}, 323
  (2004).

\bibitem{Datta90}
S.~Datta and B.~Das, {\em Appl. Phys. Lett.\/} {\bf 56}, 665 (1990).

\bibitem{Takayanagi99}
J.~Nitta, F.~E. Meijer, and H.~Takayanagi, {\em Appl. Phys. Lett.\/} {\bf 75},
  695 (1999).

\bibitem{Richter01}
D.~Frustaglia, M.~Hentschel, and K.~Richter, {\em Phys. Rev. Lett.\/} {\bf 87},
  256602 (2001).

\bibitem{D'Amico03}
R.~Ionicioiu and I.~D'Amico, {\em Phys. Rev. B\/} {\bf 67}, 041307 (2003).

\bibitem{Richter03}
M.~Popp, D.~Frustaglia, and K.~Richter, {\em Nanotechnology\/} {\bf 14}, 347
  (2003).

\bibitem{Vasilopoulos04}
B.~Molnar, F.~M. Peeters, and P.~Vasilopoulos, {\em Phys. Rev. B\/} {\bf 69},
  155335 (2004).

\bibitem{Richter04a}
M.~Hentschel, H.~Schomerus, D.~Frustaglia, and K.~Richter, {\em Phys. Rev. B\/}
  {\bf 69}, 155326 (2004).

\bibitem{Richter04b}
D.~Frustaglia, M.~Hentschel, and K.~Richter, {\em Phys. Rev. B\/} {\bf 69},
  155327 (2004).

\bibitem{Richter04c}
D.~Frustaglia and K.~Richter, {\em Phys. Rev. B\/} {\bf 69}, 235310 (2004).

\bibitem{Sigrist05}
U.~Aeberhard, K.~Wakabayashi, and M.~Sigrist, {\em Phys. Rev. B\/} {\bf 72},
  075328 (2005).

\bibitem{Vasilopoulos05}
X.~F. Wang and P.~Vasilopoulos, {\em Phys. Rev. B\/} {\bf 72}, 165336 (2005).

\bibitem{Peeters06}
P.~F{{\"o}}ldi, O.~K{{\'a}}lm{{\'a}}n, M.~G. Benedict, and F.~M. Peeters, {\em
  Phys. Rev. B\/} {\bf 73}, 155325 (2006).

\bibitem{Perroni06}
V.~M. Ramaglia, V.~C. nad G.~{De Filippis}, and C.~A. Perroni, {\em Phys. Rev.
  B\/} {\bf 73}, 155328 (2006).

\bibitem{Cohen07}
G.~Cohen, O.~Hod, and E.~Rabani, Cond. Mater. arXiv:0710.4770.

\bibitem{Levine02a}
F.~Remacle, K.~C. Beverly, J.~R. Heath, and R.~D. Levine, {\em J. Phys. Chem.
  B\/} {\bf 106}, 4116 (2002).

\bibitem{Levine02b}
K.~C. Beverly, J.~L. Sample, J.~F. Sampaio, F.~Remacle, J.~R. Heath, and R.~D.
  Levine, {\em Proc. Natl. Acc. Sci. (USA)\/} {\bf 99}, 6456 (2002).

\bibitem{Levine02c}
F.~Remacle and R.~D. Levine, {\em Israel J. Chem.\/} {\bf 42}, 269 (2002).

\bibitem{Levine03}
F.~Remacle, K.~C. Beverly, J.~R. Heath, and R.~D. Levine, {\em J. Phys. Chem.
  B\/} {\bf 107}, 13892 (2003).

\bibitem{Levine04}
F.~Remacle, I.~Willner, and R.~D. Levine, {\em J. Phys. Chem. B\/} {\bf 108},
  18129 (2004).

\bibitem{Levine05a}
F.~Remacle, J.~R. Heath, and R.~D. Levine, {\em Proc. Natl. Acc. Sci. (USA)\/}
  {\bf 102}, 5653 (2005).

\bibitem{Levine05b}
E.~Katz, R.~Baron, I.~Willner, N.~Richke, and R.~D. Levine, {\em Chem. Phys.
  Chem.\/} {\bf 6}, 2179 (2005).

\bibitem{Beenakker91}
C.~W.~J. Beenakker, {\em Phys. Rev. B\/} {\bf 44}, 1646 (1991).

\bibitem{Kinaret92}
J.~M. Kinaret, Y.~Meir, N.~S. Wingreen, P.~A. Lee, and X.~G. Wen, {\em Phys.
  Rev. B\/} {\bf 46}, 4681 (1992).

\bibitem{Bonet02}
E.~Bonet, M.~M. Deshmukh, and D.~C. Ralph, {\em Phys. Rev. B\/} {\bf 65},
  045317 (2002).

\bibitem{Hettler03}
M.~H. Hettler, W.~Wenzel, M.~R. Wegewijs, and H.~Schoeller, {\em Phys. Rev.
  Lett.\/} {\bf 90}, 076805 (2003).

\bibitem{Elste05}
F.~Elste and C.~Timm, {\em Phys. Rev. B\/} {\bf 71}, 155403 (2005).

\bibitem{Braig05}
S.~Braig and P.~W. Brouwer, {\em Phys. Rev. B\/} {\bf 71}, 195324 (2005).

\bibitem{Mukamel06}
U.~Harbola, M.~Esposito, and S.~Mukamel, {\em Phys. Rev. B\/} {\bf 74}, 235309
  (2006).

\bibitem{Hanggi2006}
F.~J. Kaiser, M.~Strass, S.~Kohler, and P.~H{\"{a}}nggi, {\em Chem. Phys.\/}
  {\bf 322}, 193 (2006).

\bibitem{Datta_book}
S.~Datta, {\em Electronic Transport in Mesoscopic Systems\/} (Cambridge
  University Press, Cambridge, 1995).

\bibitem{Gefen05}
J.~Konig and Y.~Gefen, {\em Phys. Rev. B\/} {\bf 71}, 201308 (2005).

\bibitem{Oreg05}
M.~Sindel, A.~Silva, Y.~Oreg, and J.~{von Delft}, {\em Phys. Rev. B\/} {\bf
  72}, 125316 (2005).

\end{thebibliography}
\end{document}